\documentclass[preprint2]{aastex}

\shorttitle{M84 Jet}
\shortauthors{Harris, Finoguenov, Bridle, Hardcastle, \& Laing}

\begin{document}


\title{X-ray Detection of the Inner Jet in the Radio Galaxy M84}


\author{D. E. Harris}
\affil{Smithsonian Astrophysical Observatory, 60 Garden Street,
Cambridge, MA 02138}
\email{harris@cfa.harvard.edu}

\author{A. Finoguenov}
\affil{Max-Planck-Institut f\"ur extraterrestrische Physik,
Giessenbachstra\ss e, 85748 Garching, Germany}
\email{alexis@xray.mpe.mpg.de}

\author{A. H. Bridle}
\affil{National Radio Astronomy Observatory, 520 Edgemont Road,
Charlottesville, VA 22903-2475}
\email{abridle@nrao.edu}

\author{M. J. Hardcastle}
\affil{Dept. of Physics, University of Bristol, Tyndall Avenue,
Bristol BS8 1TL, UK}
\email{M.Hardcastle@bris.ac.uk}

\and

\author{R. A. Laing}
\affil{University of Oxford, Department of Astrophysics, Nuclear and
Astrophysics Laboratory, Keble Road, Oxford OX1 3RH, U.K.}
\email{rlaing@astro.ox.ac.uk}



\begin{abstract}
During the course of an investigation of the interaction of the radio
galaxy M84 and its ambient cluster gas, we found excess X-ray emission
aligned with the northern radio jet.  The emission extends from the
X-ray core of the host galaxy as a weak bridge and then brightens to
a local peak coincident with the first detectable radio knot at
$\approx~2.5^{\prime\prime}$ from the core.  The second radio knot at
3.3$^{\prime\prime}$ is brighter in both radio and X-rays.  The X-ray
jet terminates 3.9$^{\prime\prime}$ from the core.  Although all the
evidence suggests that Doppler favoritism augments the emission of the
northern jet, it is unlikely that the excess X-ray emission is
produced by inverse Compton emission.  We find many similarities
between the M84 X-ray jet and recent jet detections from Chandra data
of low luminosity radio galaxies.  For most of these current
detections synchrotron emission is the favored explanation for the
observed X-rays.

\end{abstract}



\pagebreak
\newpage
\clearpage


\section{Introduction}

The radio galaxy M84 is a low luminosity (FRI type) radio galaxy in
the Virgo cluster.  We obtained Chandra observations in order to study
the interaction of the radio structures with the hot intra cluster
medium (ICM) and that work was reported in Finoguenov \& Jones (2001).
In this paper we report on X-ray emission detected from the inner 300
pc of the northern radio jet.

X-ray emission from radio jets presents us with the problem of
identifying the emission process but once this process is determined,
we can then obtain new constraints on physical parameters (Harris and
Krawczynski, 2002).  With the introduction of the relativistic beaming
model of Celotti (Celotti, Ghisellini, \& Chiaberge, 2001) and
Tavecchio (Tavecchio, et al. 2000), most X-ray emission from jets has
been interpreted as indicating either synchrotron emission or inverse
Compton scattering off the cosmic microwave background (CMB).  For
M84, we show that synchrotron emission is the probable process, as
has been found for a number of other FRI radio galaxies (Worrall,
Birkinshaw, \& Hardcastle, 2001; Hardcastle, Birkinshaw, \& Worrall
2001).  The implications of the detected X-ray emission are discussed
in sec.~\ref{sec:disc}.

We take the distance to the Virgo cluster to be 17 Mpc so that one
arcsec corresponds to 82 pc.  We follow the usual convention for flux
density, S$\propto~\nu^{-\alpha}$.

\section{X-ray data}\label{sec:xray}

The X-ray data were obtained with the ACIS-S detector on the Chandra
Observatory (obsid 803, 2000May19).  The exposure time was 30 ksec and
after filtering for high background a livetime of 28.7 ksec was
realized.  The central X-ray point source is coincident with the
position of the radio nucleus. Its spectrum is well described as an
absorbed ($N_H=(2.7\pm0.3)~\times~10^{21}$ cm$^{-2}$) power law
($\alpha=1.3\pm0.1$) with a corresponding luminosity of
$4\times10^{39}$ ergs~s$^{-1}$ in the 0.5--10 keV band.  Further
details of the data and basic processing can be found in Finoguenov \&
Jones (2001).

\begin{figure}
\epsscale{1.0}
\plotone{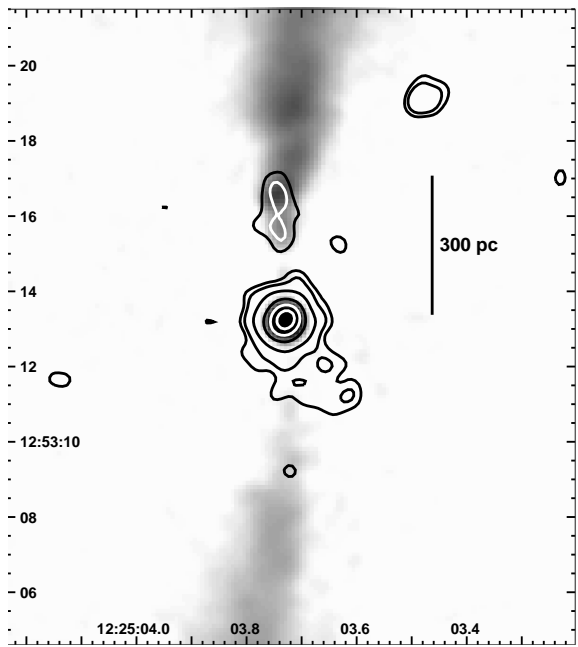}
\caption{A 5GHz VLA greyscale (restoring beam of 0.4$^{\prime\prime}$)
with X-ray contours of the inner region of M84.  The X-ray image has
been smoothed with a Gaussian of FWHM=0.5$^{\prime\prime}$ and shifted
by 0.09$^{\prime\prime}$ in RA to align the core with the radio
nucleus.  A few contours have been changed from black to white to
improve visibility.  The contour levels are 0.35, 0.5, 1, 2, 3, and 4
counts per pixel.  Since the pixel size is 0.0984$^{\prime\prime}$,
these levels should be multiplied by 103.3 to obtain counts per square
arcsec.  The X-ray detections N2.5 and N3.3 coincide with the first
two radio features in the northern jet.  The 2$\sigma$ feature
4$^{\prime\prime}$ to the south of the nucleus (a single contour)
contains 10$\pm$5 net counts.  The scale bar shows the length of 300pc
at the distance of M84.
\label{fig:over}}

\end{figure}

For image display purposes, we limited the energy range to 0.3-2keV
and binned the data by a factor of 1/5 to obtain images with pixel
size 0.098$^{\prime\prime}$.  Various Gaussian smoothing functions
were then convolved with the data and one example is shown in
figure~\ref{fig:over}; a radio image with X-ray contours overlayed.
The coincidence of radio and X-ray emissions for knots N2.5 and N3.3
is evident (we name these features according to their angular distance
from the core, in arcsec).  The emission to the SW of the core at
about the same brightness level as the northern jet does not
correspond to any radio emission and is part of the complex thermal
emission described by Finoguenov \& Jones (2001).  The data presented
there revealed an H-shape of the diffuse component, anticorrelated
with radio emission. Furthermore, the asymmetry in the X-ray emission,
as well as in the radio, is attributed to hydrodynamic effects caused
by motion of M84 towards the North-West through the hot gas of the
Virgo cluster.  As is clearly seen from their large-scale map, the
western component of the X-ray emission is more compressed than the
Eastern one and in addition is bent towards the south.  The SW
enhancement in the X-ray emission, seen in our fig.~\ref{fig:over}, is
attributed to the stronger interaction between the ambient hot plasma
and the southern radio lobe on the side of maximal compression of the
X-ray emitting plasma.

\normalsize

For an estimate of the spectral parameters, a thermal (MEKAL) spectrum
was fit to a large region near the north jet.  Then with the counts
extracted from a circle of radius 2.5$^{\prime\prime}$ which included
both N2.5 and N3.3, a two component model was fit with the temperature
and metalicity of the thermal component fixed to that found for the
adjacent hot gas (0.55$\pm0.05$~keV), but with the amplitudes of both
the power law and the thermal components left as free parameters (as
was the power law index).  N$_H$ is always fixed to the Galactic
value.

A single component fit to the jet spectrum, allowing only the
normalization to vary gives an unacceptable $\chi^2$: 51.3 for 16
degrees of freedom.  The introduction of an additional component, such
as the power-law, reduces the $\chi^2$ to an acceptable level of 13.5 per
14 degrees of freedom and the additional component is statistically
significant at the 3.9 $\sigma$ level (or 99.992\%).  If the second
component is modeled with a thermal spectrum instead of a power law
and only the temperature and normalization are allowed to vary, the
fit is equally acceptable (13.9/14) and has a temperature of
3.2$\pm$1.1~keV and normalization of (8.7$\pm$3.3)~$\times~10^{-6}$ in
XSPEC units.

\begin{figure}
\plotone{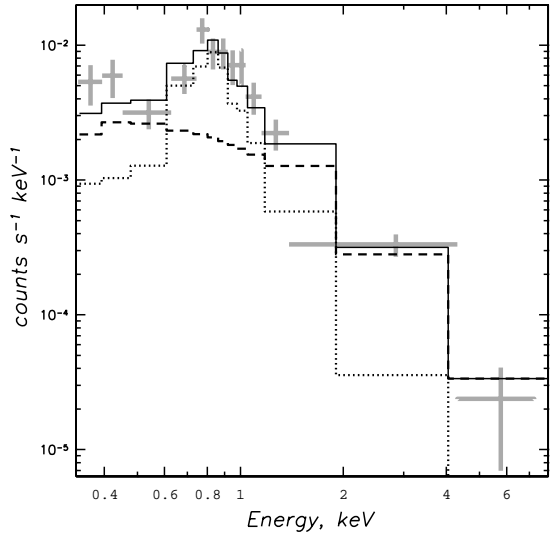}
\caption{The X-ray spectrum of emission extracted using the
2.5$^{\prime\prime}$ diameter circle, centered on the north X-ray jet.
Grey crosses indicate the Chandra data; the dotted line is
the contribution of the thermal component from diffuse emission of M84;
the dashed line indicates the power law spectrum of the jet; and the
solid line corresponds to the sum of the two
components.\label{fig:xspec}}
\end{figure}

The results for the thermal background/power-law-jet are shown in
fig.~\ref{fig:xspec}.  From this fit, we find that the power law
component has a flux,
f(0.3-7keV)~=~1.5$\times~10^{-14}$~erg~cm$^{-2}$~s$^{-1}$ and
$\alpha=0.8~\pm~0.3$. This flux corresponds to 42 net counts in our
data.  Dividing the observed flux according to the ratio of the counts
in two small circles centered on the two knots provides the intensity
values reported in table~\ref{tab:inten}.

\begin{deluxetable}{lccl}
\tabletypesize{\scriptsize}
\tablecaption{Intensity Measurements for the core and knots of M84. \label{tab:inten}}
\tablewidth{0pt}
\tablehead{
\colhead{Parameter} & \colhead{N2.5}   & \colhead{N3.3} & \colhead{units}\\
}
\startdata
net counts\tablenotemark{a} & 15$\pm$5 & 28$\pm$6 & counts\\
f$_x$(0.3-7keV)\tablenotemark{b}  & 0.53$\pm$0.18 & 0.96$\pm$0.20 &10$^{-14}$~erg~s$^{-1}$~cm$^{-2}$\\
L$_x$(0.3-7keV) &  1.8$\pm$0.6 & 3.3$\pm$0.7 & 10$^{38}$~erg~s$^{-1}$ \\
S(1keV) &  0.63$\pm$0.21 & 1.15$\pm$0.25 & nJy \\
S(5GHz)\tablenotemark{c} &  3.5$\pm$0.6 & 13$\pm$3 & mJy \\
\enddata


\tablenotetext{a}{The X-ray intensity of the knots comes from the
spectral modeling, but the listed counts come from small circles:
r=0.35$^{\prime\prime}$ for N2.5 and r=0.5$^{\prime\prime}$ for N3.3.
This was done because the two knots are not well separated, being only
0.8$^{\prime\prime}$ apart.  Obviously the net counts in these small
circles do not represent the total counts attributable to the knots,
but their sum is close to the 42 counts contributing to the power law
component of the spectral fit within a circle of radius
2.5$^{\prime\prime}$.  Listed uncertainties are the square root of the
total counts within the measuring aperture, but they should be
augmented by a few counts because of the uncertainty in the
appropriate background level to use at the location of the knots.}

\tablenotetext{b}{The X-ray flux has been corrected for galactic
absorption and a power law with $\alpha$=0.8 was used for both components.}

\tablenotetext{c}{The radio flux densities were measured with a variety
of AIPS tools.}

\end{deluxetable}

\normalsize

\section{Radio data}\label{sec:radio}

The VLA data are a combination of observations made in three VLA
configurations at 4.9 GHz: in the A configuration on 1980 November 9,
in the B configuration on 1981 June 25, and in the C configuration on
1981 November 17. They were reduced in the AIPS software package using
standard self-calibration and imaging methods.  The asymmetry in
brightness between the northern and southern jets supports the notion
that the northern jet is the one coming towards us and that Doppler
favoritism is operating out to $\approx~15^{\prime\prime}$ (1.2~kpc).

We do not have radio data of sufficient spatial resolution at lower
frequencies to determine accurate spectral indices, but the lower
resolution data indicate values $\alpha~\approx$~0.65.

\section{Parameters for Emission Models}

To estimate physical parameters associated with various X-ray emission
mechanisms, we need to assign volumes for the knots.  For each knot we
choose a cylindrical volume with length of 0.8$^{\prime\prime}$ and
radii of 0.15$^{\prime\prime}$ (N2.5) and 0.21$^{\prime\prime}$
(N3.3).

\begin{figure}
\plotone{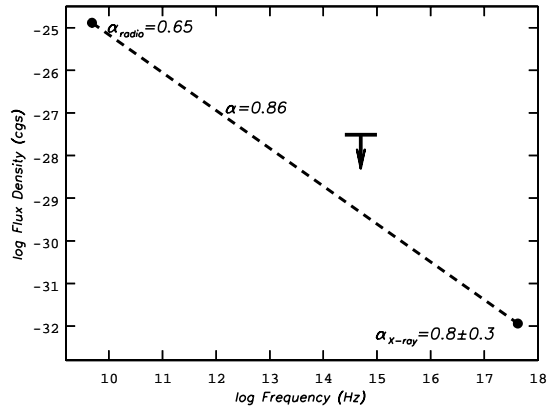}
\caption{The spectrum of knot N3.3 (knot N2.5 is quite similar).  The
optical upper limit is from archival HST data: 30~$\mu$Jy at
5~$\times~10^{14}$~Hz.  The other two points are from
Table~\ref{tab:inten}.\label{fig:spec}}
\end{figure}

The overall spectrum for N3.3 is shown in figure~\ref{fig:spec}.

\subsection{Thermal Bremsstrahlung Emission}

In many instances (e.g. Harris, Carilli, \& Perley, 1994) it has been
argued that if X-ray emission from radio features were to be from hot
gas rather than a non-thermal mechanism, then the required electron
densities together with the (equipartition) magnetic field strengths
would predict departures from the $\lambda^2$ relation of radio
polarization position angle, as well as excess depolarization.  Since
these effects have not been found, it appears that
whatever thermal gas is present in radio emitting volumes must have a
density much lower than that needed to produce the observed X-ray
emission.  However, there could be excess hot gas around the jet since
we do not have sufficient resolution or s/n to determine the precise
distribution of the emission and we do not have Faraday rotation
estimates on the arcsec scales under discussion here.

The X-ray luminosity for each knot is of order 10$^{38}$~erg~s$^{-1}$
(table~\ref{tab:inten}) and the density required to produce this
emission from the cylindrical volumes would be 5~cm$^{-3}$.  The
masses of the emitting cylinders would be 4 and
7~$\times~10^3~M_{\odot}$ (N2.5 and N3.3, respectively).  Assuming a
temperature of 1 keV means that the pressure would be
1.6$~\times~10^{-8}$dyne~cm$^{-2}$.  Although this pressure is
significantly larger than that expected from the ambient hot gas
(3.7$~\times~10^{-10}$dyne~cm$^{-2}$, Finoguenov \& Jones, 2001) ,
since these emitting volumes are well inside the galaxy, there could
be additional pressure contributed by cooler gas which does not
produce X-rays.

The X-ray spectral analysis (section~\ref{sec:xray}) demonstrates
clear evidence for either a non-thermal or a 3~keV component not seen
in nearby regions.  In figure~\ref{fig:xspec}, note the excess both
below 0.5~keV and above 2 keV.  We favor the non-thermal alternative
because even adiabatic compression of the X-ray emitting gas to the
required densities ($\geq$~1~cm$^{-3}$) should result in temperature of
15 keV, much higher than what is observed.

\subsection{Synchrotron Emission}

Even though we do not know the details of the spectrum, we may
estimate the synchrotron parameters necessary to produce the observed
X-rays with a spectrum such as that shown in fig.~\ref{fig:spec}.  For
the radio emission from N3.3 (10$^7$ to 10$^{11}$~Hz), the log of the
luminosity would be 38.24~erg~s$^{-1}$ and the equipartition field
would be 111~$\mu$G.  For a synchrotron X-ray model, we need to extend
the spectrum up to 10$^{18}$~Hz with the spectral index $\alpha$=0.9.
In this case the log of the luminosity would be 39.14~erg~s$^{-1}$
and the equipartition field would be 143~$\mu$G.  If the synchrotron 
emission from N2.5 and N3.3 is mildly beamed (as we argue below) both 
these luminosity and magnetic field strength estimates should be
reduced somewhat.

Although the change required by the extension of the synchrotron
spectrum in the total energy contained in the source is not large, the
power law distribution of electron energy would have to extend to
$\gamma=4~\times~10^7$ with a halflife of 10 years for electrons of
this energy.  If $\alpha$(radio)=0.65, then the spectrum would most
likely be a broken power law, but this will not make much difference
to the rough estimates calculated here.

\subsection{Inverse Compton Emission}

The synchrotron self-Compton model fails because the photon energy
density is so low that the predicted 1 keV flux density would be 4
orders of magnitude below that observed (assuming an equipartition
field of 93 and 111~$\mu$G for N2.5 and N3.3, respectively).

IC scattering off the CMB photons would require a magnetic field
strength of 0.1~$\mu$G, more than a factor of 1000 below the
equipartition field.

If we invoke relativistic beaming to produce the observed X-ray jet
(so that the CMB energy density would be increased by $\Gamma^2$ in
the jet frame; where $\Gamma$ is the Lorentz factor of the
relativistic jet), an angle between the jet velocity vector and the
line of sight would have to be 3$^{\circ}$ or less and the beaming
factor and $\Gamma$ would be $\approx$~20.  These values, derived from
the equations in the appendix of Harris \& Krawczynski (2002), are
inconsistent with estimates from the radio data.  From the observed
ratio of intensities of the inner radio jets (i.e. 16 at
3$^{\prime\prime}$ from the core), the angle between the line of sight
and the N jet has to be less than 70$^{\circ}$ and is most likely
greater than 45$^{\circ}$ since we see the two sides of the jet and
lobes well separated.  This range in angles corresponds to beaming
factors in the range 0.6 to 1.25 and jet fluid velocities,
$\beta~(=~\frac{v}{c}$), in the range 1 to 0.6.

\section{Discussion}\label{sec:disc}

We believe the evidence favors synchrotron emission for the observed
X-rays although to sustain this model, higher s/n X-ray data and
optical detections are required.  Undoubtedly there are bulk
relativistic velocities in the jet producing the observed intensity
differences between the N and S jets, but with velocity vectors not
too far from the plane of the sky, we see only mild boosting and the
parameters for IC/CMB emission are completely at odds with all other
evidence.  To check the 'mild beaming' synchrotron model, we note that
the ratio of the net counts in the N jet (r=1.05$^{\prime\prime}$
aperture) to that found in the same sized circle at the same distance
to the south  is $>$4.1 (where we used the 2$\sigma$ upper limit for
the south value).  Thus the X-ray ratio (North/South) is consistent
with the radio value (16) measured at the same distance from the core
(3$^{\prime\prime}$). 

  Even in fields of order 100 $\mu$G, the half-life for electrons
producing X-rays is so short that they could travel no more than
$\approx$~3pc from their acceleration region.  Thus the X-rays clearly
demarcate that sort of acceleration region.

The M84 jet is considerably weaker than other FRI detections.  The
X-ray luminosity is $\approx$~6 times less than that of Cen A, the jet
with the lowest luminosity listed in a table of 7 FRI jets in Harris,
Krawczynski, \& Taylor (2002).  The combined flux densities of N2.5 and
N3.3 are a factor of 84 less than that from knots HST-1 and D in the
M87 jet (Marshall et al. 2002); these two knots are within
3$^{\prime\prime}$ of the nucleus of M87 although their physical
distance from the core may be larger than that for the M84 knots owing
to a larger projection effect.  Although some of the disparity in
X-ray luminosity between M84 and the other FRI jets may be caused by
differences in beaming factors, most of the current sample are
believed to have rather large angles between the line of sight and the
jet axis and thus beaming is generally moderate to low in all of them.



\acknowledgments

We thank the referee for suggestions which led to improvements in the
presentation.  This work was partially supported by NASA grants and
contracts GO0-1145X, GO1-2135A and NAS8-39073.  The National Radio
Astronomy Observatory is a facility of the National Science Foundation
operated under cooperative agreement by Associated Universities, Inc.


References\\



\noindent
Hardcastle, M.J., Birkinshaw, M., \& Worrall, D.M. 2001 \mnras\ 326, 1499 3c66B

\noindent
Hardcastle, M.J., Worrall, D.M., Birkinshaw, M., Laing, R.A. \& Bridle,
A.H. 2002, \mnras\ (in press)
3C31

\noindent
Harris, D. E., Krawczynski, H., and Taylor, G.B. 2002 ApJ (in press)

\noindent
Harris,  D. E. and Krawczynski, H. 2002 \apj\ 565, 244

\noindent
Harris, D.E., Carilli, C.L. and Perley, R.A. 1994 Nature 367, 713.

\noindent
Finoguenov, A. \& Jones, C. 2001 \apj\ 547, L107


\noindent
Marshall, H.L., Miller, B.P., Davis, D.S., Perlman, E.S., Wise, M.,
Canizares, C.R., and Harris, D.E. 2002, ApJ 564, 683

\noindent
Worrall, D.M., Birkinshaw, M., \& Hardcastle, M.J. 2001, \mnras\ 326, L7



\end{document}